\documentclass[
  journal=pasa,
  manuscript=research-paper, 
  year=202X,
  volume=Y,
]{cup-journal}

\usepackage{microtype,siunitx,booktabs}
\usepackage{amsmath, amssymb, natbib}
\usepackage{subcaption}
\usepackage{float}

\usepackage[inkscapearea=page]{svg}
\usepackage{adjustbox}

\newcommand{\lya}{Lyman-$\alpha$}
\newcommand{\oiii}{[\hbox{{\rm O}\kern 0.1em{\sc iii}}]\,5007}
\newcommand{\oii}{[\hbox{{\rm O}\kern 0.1em{\sc ii}}]\,3727}
\newcommand{\ha}{\hbox{{\rm H}\kern 0.1em{\sc $\alpha$}}}
\newcommand{\hb}{\hbox{{\rm H}\kern 0.1em{\sc $\beta$}}}

\title{MOSEL survey: Unwrapping the Epoch of Reionization through mimic galaxies at Cosmic Noon}

\author{Ravi Jaiswar}
\affiliation{International Centre for Radio Astronomy Research, Curtin University, Bentley WA, Australia}
\alsoaffiliation{ARC Centre of Excellence for All Sky Astrophysics in 3 Dimensions, Bentley WA, Australia}

\author{Anshu Gupta}
\affiliation{International Centre for Radio Astronomy Research, Curtin University, Bentley WA, Australia}
\alsoaffiliation{ARC Centre of Excellence for All Sky Astrophysics in 3 Dimensions, Bentley WA, Australia}

\author{Elisabete da Cunha}
\affiliation{International Centre for Radio Astronomy Research, University of Western Australia, Crawley WA, Australia}
\alsoaffiliation{ARC Centre of Excellence for All Sky Astrophysics in 3 Dimensions, Bentley WA, Australia}

\author{Cathryn M. Trott}
\affiliation{International Centre for Radio Astronomy Research, Curtin University, Bentley WA, Australia}
\alsoaffiliation{ARC Centre of Excellence for All Sky Astrophysics in 3 Dimensions, Bentley WA, Australia}

\author{Anishya Harshan}
\affiliation{University of Ljubljana, Department of Mathematics and Physics, Jadranska ulica 19, SI-1000 Ljubljana, Slovenia}
\alsoaffiliation{ARC Centre of Excellence for All Sky Astrophysics in 3 Dimensions, Bentley WA, Australia}

\author{Andrew Battisti}
\affiliation{Research School of Astronomy and Astrophysics, Australian National University, Canberra, ACT 2611, Australia}
\alsoaffiliation{ARC Centre of Excellence for All Sky Astrophysics in 3 Dimensions, Bentley WA, Australia}

\author{Ben Forrest}
\affiliation{Department of Physics and Astronomy, University of California Davis, One Shields Avenue, Davis, CA, 95616, USA.}

\doi{ZZZ}

\received {dd Mmm YYYY}
\revised  {dd Mmm YYYY}
\accepted {dd Mmm YYYY}
\published{dd Mmm YYYY}


\begin{document}

\begin{abstract}\label{sec:abstract}
The nature of the first galaxies that reionized the universe during the Epoch of Reionization (EoR) remains unclear. Attempts to directly determine spectral properties of these early galaxies are {affected by both limited photometric constraints across the spectrum} and by the opacity of the intergalactic medium (IGM) to the Lyman Continuum (LyC) at high redshift. We approach this by analysing properties of {analogous} extreme emission line galaxies (EELGs, [OIII]+Hbeta EW $>400 \AA$) at $2.5<z<4$ from the ZFOURGE survey using the Multi-wavelength Analysis of Galaxy Physical Properties (MAGPHYS) SED fitting code. We compare these to galaxies at $z>5.5$ observed with the James Webb Space Telesope (JWST) with self-consistent spectral energy distribution fitting methodology. This work focuses on the comparison of their UV slopes ($\beta_P$), ionizing photon production efficiencies $\xi_{ion}$, {star formation {rates} and dust properties} to determine the effectiveness of this {analogue} selection technique. We report the median ionizing photon production efficiencies as log$_{10}(\xi_{ion}/(Hz\  {\rm erg}^{-1}))=$$25.14^{+0.06}_{-0.04}$,$25.16^{+0.06}_{-0.05}$,$25.16^{+0.04}_{-0.05}$,$25.18^{+0.06}_{-0.07}$ for our {ZFOURGE control, ZFOURGE EELG}, JADES and CEERS {samples} respectively. {ZFOURGE }EELGs are 0.57 dex lower in stellar mass and have half the dust extinction, compared to their ZFOURGE control counterparts. They also have a similar specific star formation rates and $\beta_P$ to the $z>5.5$ samples. {We find that EELGs at low redshift ($2.5<z<4$) are analogous to EoR galaxies {in their dust attenuation and specific star formation rates}. {T}heir extensive photometric coverage and the accessibility of their LyC region opens pathways to infer stellar population properties in the EoR.}  

\end{abstract}

\section{Introduction} \label{sec:intro}
The model of the Universe's evolution through the Epoch of Reionization is {directly affected by} our limited understanding of the first galaxies. Debate of their ionizing capabilities persists due to their relatively unconstrained ionizing photon production and escape fractions \citep{Robertson2010,Madau2015,Naidu2019,Finkelstein2019}. While AGN-quasars are unlikely {ionization source} candidates due to their infrequency beyond $z>3$ \citep{Kulkarni2018}, bright, highly star-forming galaxies could have produced copious ionizing photons for reionization \citep{Naidu2019}. Alternatively, numerous faint galaxies could provide sufficient ionizing flux over a longer history to reionize the intergalactic medium \citep{Bruton2023}.

Operation of the {James Webb Space Telescope }(JWST) has, in a short time, revolutionized our understanding of galaxies in the Epoch of Reionization (EoR) and this landscape is constantly changing with new revelations \citep{Bunker2023, Bagley2023,Paris2023}. {Early empirical studies had noticed potentially large \oiii\+\hb\ emissions at higher redshift} \citep{Schaerer2009,Raiter2010}. {Now with recent redshift evolution models} \citep{Zhai2019} {and JWST/NIRCam number density studies $z>5.3$ }\citep{Matthee2023}{ this is evidently ubiquitous in the early universe}. Probing galaxies within the EoR reveals almost no dust attenuation {as expected}, particularly approaching the highest redshift limits \citep{Robertson2023,Hsiao2023,Tacchella2023,Haro2023}. {However, galaxies with} unexpectedly high stellar masses have been found \citep{Boylan-Kolchin2022,Labbe2022} bordering the possible limits set by $\Lambda$CDM and suggesting extremely efficient star formation. Exceptional findings such as a highly quenched, dusty galaxy at $z\sim5$ \citep{Donnan2022,Harikane2022,Naidu2022,Haro2023b,Yung2023} and spectroscopic confirmation of a $z\sim13.2$ \citep{Robertson2022} galaxy have {both} challenged and affirmed ideas of the timescale over which the Universe evolves.

\begin{figure*}[t]
    \centering
    \includegraphics[width=0.95\textwidth, height=0.33\textheight]{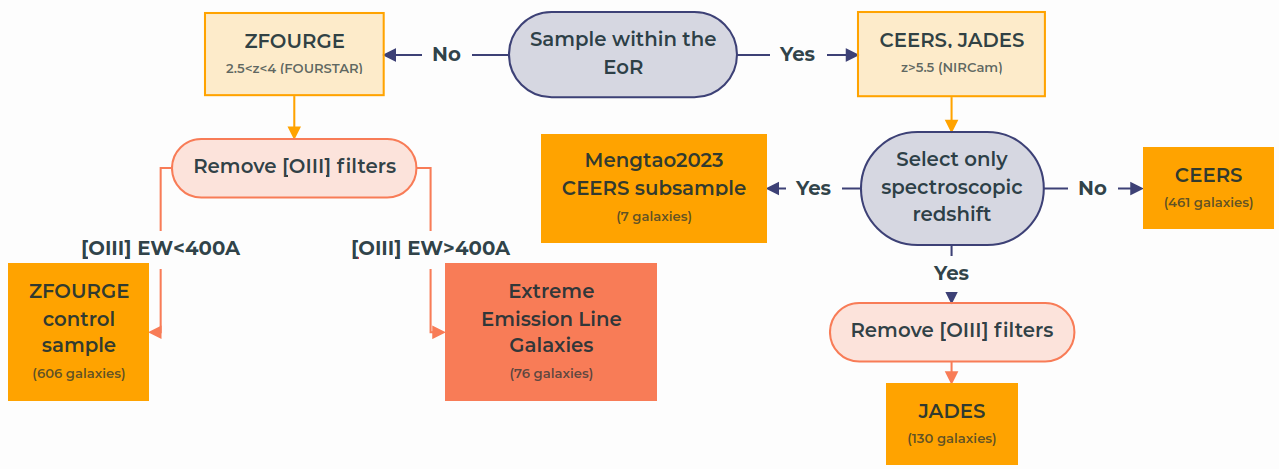}
    \caption{Summary of samples and their constraining quantities as well as some processing information}
    \label{fig:flowchart}
\end{figure*}
\begin{figure*}
    \centering
    \includegraphics[width=0.95\textwidth, height=0.40\textheight]{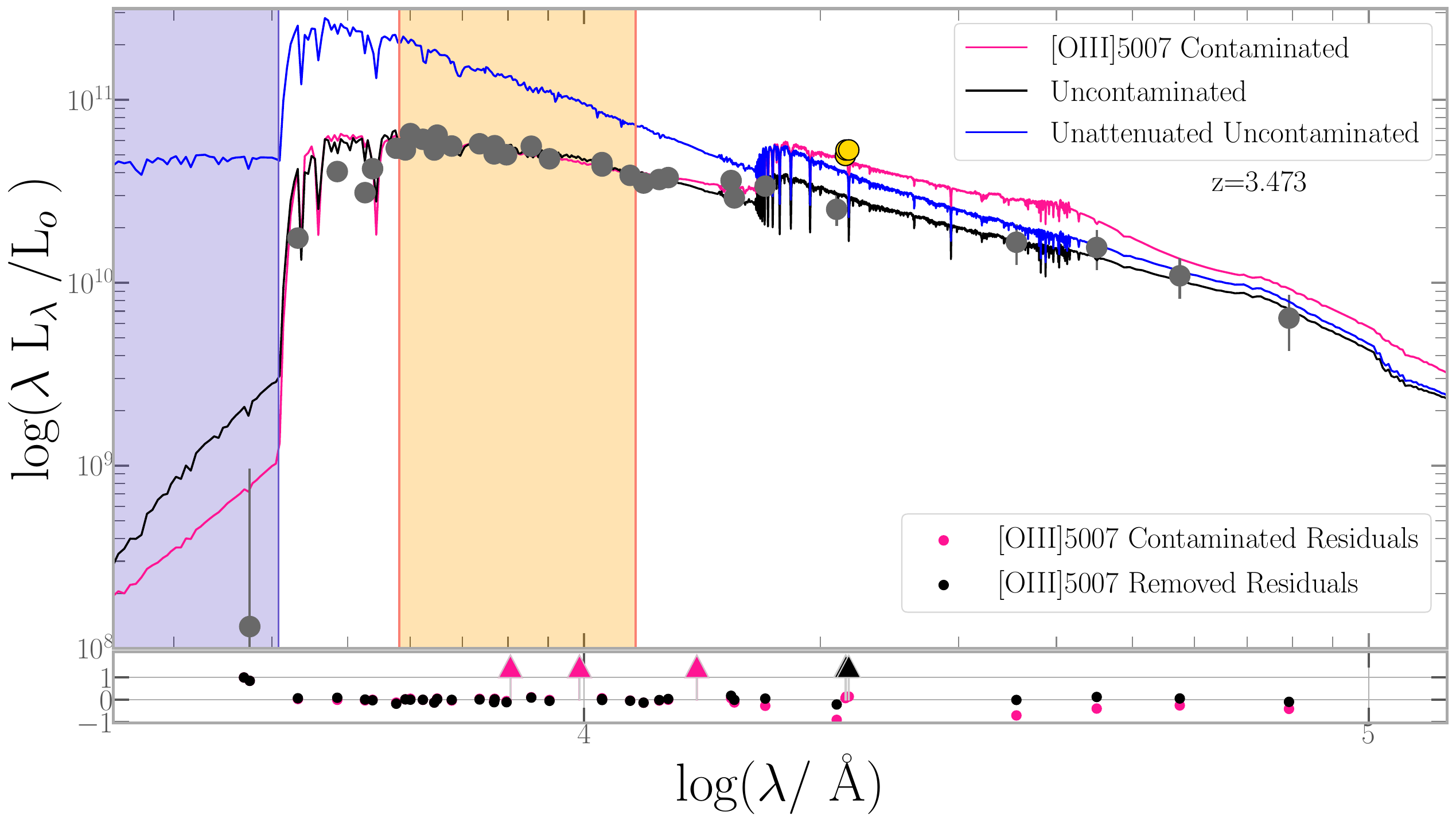}
    \caption{Observed frame MAGPHYS SED of z=3.473 {ZFOURGE }EELG indicating the models fit to the photometric filter fluxes (grey circles) and their residuals (black circles, arrows if residual $\sigma_{res}>1$). 
    Blue region represents the LyC and is integrated under the unattenuated uncontaminated line (blue) to derive log$_{10}(\xi_{ion}/(Hz\  {\rm erg}^{-1}))$ photometrically. Orange region is the UV region from which the UV slope is determined from either the attenuated SED $\beta_a$ or from the photometry $\beta_p$. Gold filters represent the \oiii\ contaminated filters and the difference these make to the attenuated SED is shown in the continuum luminosity difference between the `uncontaminated' (black) and `contaminated' (pink) lines. Residuals of the contaminated bestfit are represented by pink circles where $\sigma_{res} > 1$ are represented by pink arrows. }
    \label{fig:regionplot}
\end{figure*} 

{The LyC is the component of stellar emission with sufficient energy to ionize neutral hydrogen atoms and is a key parameter in theoretical models of EoR.} However, deriving physical properties relating to the production of LyC radiation and their escape from the interstellar medium of galaxies comes with both technical and physical challenges. Direct detection of LyC photons is impeded by the opacity of neutral hydrogen in EoR galaxies, instead requiring models based on lower redshift analogues \citep{Izotov2018}. LyC leakers at low redshift exhibit high \oiii\ EWs and high O$_{32}$ (\oiii/\oii\ ) ratios albeit with large scatter \citep{Cardamone2009,Izotov2018}. For galaxies ${\rm z}<0.5$, a \lya\ peak separation of less than 200\,km/s is the strongest predictor of significant LyC escape at ${\rm z}\sim0$ ($f_{esc}>0.1$ up to 0.8) \citep{Izotov2022}. This is the basis of EoR-\lya\ correlated $f_{esc}$ predictions, with $>10$\, Gyr between the two epochs. 

It is possible that these low redshift $f_{esc}$ probes may have a different relationship with LyC escape at high redshifts due to alternate escape methods. \cite{Ji2020} find no \lya\ emission (only absorption) despite strong LyC detections in a $z\sim3.8$ galaxy; favouring a model of $f_{esc}$ through LyC-transparent holes rather than through an optically thin interstellar medium (ISM). The assumption that $z<0.5$ and $z>5$ galaxies are comparable is in contention with our understanding of these epochs \citep{Madau2014}. Studies using one epoch to analyse another must contend with {their fundamental differences and} what their comparisons really reveal. For example, the \lya\ probe or UV slope may have a different {relative} dependence on dust extinction and the stellar population {due to morphological differences \citep{Hayes2011,Meng2020}}.

{With the advent of the JWST, recent observations targeting the \ha\ and \hb\ Balmer lines have been used to estimate the production efficiency of hydrogen ionizing photons ($\xi_{ion}$) for many $z>6$ sources \citep{Tang2023,Simmonds2023}. However, with the existing uncertainty in dust content of high redshift sources, emission line corrections may not accurately reproduce the original luminosity as intended. Furthermore, limited spectroscopic availability means statistically significant {studies} come to rely on photoionization models \citep{Ferland1998,Levesque2010} to estimate $\xi_{ion}$ \citep{Bouwens2015} and compare its correlation to other ISM probes such as \oiii+\hb EW \citep{Kewley2015,Tang2019,Tang2023}}

{The lack of {photometric measurements/coverage} between the UV and NIR diminishes constraints on the modelled physical properties at ${\rm z}>6$. Quantities such as metal abundances \citep{Vincenzo2018,Torrey2018,Berg2019}, the escape fraction $f_{esc}$ or the number of ionizing photons $\dot{n}$ \citep{Anderson2017,Iyer2017} rely heavily on star formation history (SFH) models to {infer} the {absorbed light}, which depend{s} strongly on {properly sampling the spectrum}. }


{Having a well sampled SED for galaxies with well constrained redshifts allows for the determination of a photometric UV slope ($\beta_p$). This probes the empirical emission ``blueness" which is used in dust corrections \citep{Calzetti1994,Reddy2018} and as a proxy for the stellar ages and metal/dust content. Studies of local universe starbursts find that galaxies with {a bluer, more} negative model $\beta$ (determined from $f_\lambda$) have lower dust obscuration, similar to $z\sim2$ studies \citep{Takeuchi2012,Sklias2013}. The relationship of $\beta$ and dust {attenuation} is degenerate with metallicity and star formation history \citep{Bouwens2016} which complicates the relative contributions of each component. However, correlations to dust content are robust at fixed UV luminosity. \cite{Reddy2018} and \cite{Nanayakkara2022} find that galaxies at $4<z<7$ are consistently blue and have very low dust attenuation.}  

The approach this paper takes to these issues in studying the potential influence of the brightest emitters {on the EoR} is to {use an established analogue} sample with the {stellar and \oiii\ emission }properties of higher redshift galaxies at the slightly lower {redshift range of }$2.5<{\rm z}<4$ (11.1-12.2 Gyr ago). The selected redshift range is temporally closer to the EoR (12.7 Gyr ago) than many low redshift comparisons, while still having an accessible LyC region for follow up observational $f_{esc}$ constraints. The sample is taken from the FourStar Galaxy Evolution Survey (ZFOURGE) \citep{Straatman2016}, which combines deep imaging with medium and narrow-band near-IR filters with multi-wavelength observations from several public surveys to accurately determine the redshifts of $\sim 70000$ objects to $\sigma_z\sim 2\% $ accuracy \citep{Nanayakkara2016,Tran2020}. 

We will be exploring a subsample of these galaxies presented by \cite{Forrest2018} as having extreme \hb+\oiii\ EW ($>400\AA$) {and therefore considered analogues of EoR galaxies \citep{Tang2023}}. Follow up spectroscopy with the KMOS /VLT {as part of the} Mutli-Object Spectroscopic Emission Line (MOSEL) survey confirms the photometric selection for {a subset of these as} EELGs \citep{Tran2020,Gupta2022}. Of 19 galaxies targeted, 16 had bright emission lines, where 14 of these had \oiii\ as the brightest line and two had \ha\,. This sample (40 filters) will be compared to direct EoR galaxy measurements made by the CEERS (14 filters) \citep{Yang2020,Bagley2023,Yang2023} and JADES (23 filters) \citep{Bunker2023,Eisenstein2023,Hainline2023,Rieke2023} surveys using legacy HST and the recent JWST observations to determine the overlap in their observed and the SED model-derived properties. For consistency, we use the Multi-wavelength Analysis of Galaxy Physical Properties (MAGPHYS) SED fitting code \citep{DaCunha2008,DaCunha2015} across all samples. 


This work is formatted so that Section \ref{sec:data} describes the data and selection philosophies used to identify Extreme Emission galaxy analogues while Section \ref{sec:methods} delves into our methods of analysis and the quantities we use to compare the samples. Results of how representative the analogues are and what their high energy emission properties look like are presented in Section \ref{sec:results}. We summarize the work in Section \ref{sec:summary}.

Throughout this paper we assume a flat universe with $\Omega_m$=0.3, $\Omega_\lambda$=0.7 and H$_0$=70km/s/Mpc for our models. 
\section{Data} \label{sec:data}

\subsection{The analogues}
{
In this section we break down the selection criteria and processing of the $2.5<z<4$ {ZFOURGE }control sample \ref{subsec:zfourge} as well as the EELG analogues \ref{subsec:eelgs}. See Fig \ref{fig:flowchart} for the sample processing.}
\subsubsection{ZFOURGE {control}}\label{subsec:zfourge}
ZFOURGE is a survey combining legacy photometric UV to NIR data from 3 well studied regions (CDFS, COSMOS and UDS) with the FOURSTAR instrument upon the Magellan telescope \citep{Straatman2016}; its J, H and K medium band filters spanning the 1-1.8$\mu {\rm m}$ range. It creates a photometrically well sampled survey and {with the inclusion of two  narrowband filters to optimally constrain the 4000\AA\ break,} enables robust 2\% accuracy redshifts for over 70,000 galaxies \citep{Nanayakkara2016}. 
In this work the ZFOURGE survey data was reduced to only retain CDFS sources as this field had the best narrowband filter depth (NB118 and NB209) which is important in isolating the \oiii\, line flux for $z\sim3$ sources. Galaxies were selected so that their K-band SNR$>$20 and were between the $2.5-4$ redshift range. {We used the} use=1 flag which eliminated catastrophic FAST \citep{Kriek2009} and EAZY \citep{Brammer2008} SED fits, interloper stars, AGN and sources too close to bright stars. The errors for IRAC bands were set to {a floor of} 25\% (if not already above this) as {the large PSF requires that the errors be large at high redshift. \citep{Straatman2016}}. These were {otherwise} unphysically small and were found to significantly reduce the number of galaxies with well constrained physical stellar parameters. This created a total sample of 682 galaxies which includes both the control sample (606 galaxies) and the Extreme Emission Line Galaxy sample (76 galaxies) (see section \ref{subsec:eelgs}).

\subsubsection{{ZFOURGE} EELGs at z$\sim$ 3}\label{subsec:eelgs}
{ZFOURGE }EELGs were identified in \cite{Forrest2018} using a stacked superposition of similar galaxy SEDs. The EELGs were determined as having an \oiii\ EW of at least 400\AA\ and were selected between $2.5<{\rm z}<4$. This creates a subsample of 76 EELGs among the 682 total galaxies within the CDFS region, described in the subsection \ref{subsec:zfourge}. {We also use 18 of the 19 reliable spectroscopic redshifts from the MOSEL survey subsample which replace the corresponding ZFOURGE {photometric} redshifts.}

\subsection{Epoch of Reionization Sample}
{In this section we look at the selection criteria and processing of the $z>5.5$ control samples. This is broken down into the photometric redshift sample taken from the CEERS survey (section \ref{subsec:ceers}) as well as the subsample of this which have spectroscopic redshifts. This is followed by the JADES survey sample which is only has spectroscopic redshifts (section \ref{subsec:jades}).}

\subsubsection{CEERS}\label{subsec:ceers}
The Cosmic Evolution Early Release Science Survey \citep{Yang2020,Bagley2023,Yang2023} combines JWST NIRCAM photometry with legacy Extended Groth Strip (EGS) field HST data with the science goal of scouting the emergence of galaxies at cosmic dawn. {We} use the September 2022 data release {processed with} Grizli\footnote{https://s3.amazonaws.com/grizli-v2/JwstMosaics/v4/index.html} {where photometric redshifts were derived with EAZY-py}. {We select galaxies within} $5.5<{\rm z}<14$ and eliminate sources with a 95\% confidence interval of the $\chi^2$ value above 0.2 Gyr ($t_{z[97.5]}-t_{z[2.5]} <0.2$Gyr), which similarly constrains the low and high ends of the redshift bounds. We further require that the determination of the UV slope be based on at least 3 filters in the UV window \citep{Calzetti1994} and remove sources which do not fit this criteria. This selects only the galaxies with a well constrained photometric redshifts and UV regions, giving us a sample of 461 galaxies within the EoR to compare to the ZFOURGE galaxies, particularly the EELG subsample.   
\subsubsection{Spectroscopic redshifts}\label{subsec:tao}
For a small subsample of the CEERS catalogue we have collected published spectroscopic redshifts from \cite{Tang2023} to minimize fitting errors within MAGPHYS and compare with the results of the publication. These will be referred to as the Tang23 sample throughout the paper.

\subsubsection{JADES}\label{subsec:jades}
The JWST Advanced Deep Extragalactic Survey \citep[JADES,][]{Bunker2023,Eisenstein2023,Hainline2023,Rieke2023} combines JWST NIRCAM photometry with legacy HST Ultra Deep Field (UDF) data. We only select spectroscopically confirmed galaxies {at }${\rm z}>5.5$, deriving a sample of 130 galaxies between $5.5<{\rm z}<13.2$. The JADES datasets depth and additional filters help to better constrain the physical parameters, and thus we compare mostly to this sample.  

\section{Methods}\label{sec:methods}
In this section we discuss the derivation of parameters used to describe the `blueness' ({UV slope}, $\beta$) of a galaxy's emission profile and its likelihood to contribute LyC to the IGM from both photometric data. {We describe the SED models used, their limitations and any} modifications we have made.   

\subsection{MAGPHYS}\label{subsec:magphys}
Multi-wavelength Analysis of Galaxy Physical Properties (MAGPHYS) is an SED fitting package which derives the physical properties of a galaxy from the supplied photometry \\ \citep{DaCunha2008,DaCunha2015}. It does this by assembling a library of dust and stellar models at a predetermined redshift (if not allowed to vary) and then approaches the closest model using a marginalized likelihood distribution of each physical parameter (for more information see the documentation of MAGPHYS). We use the BC03 stellar models \citep{Bruzual2003} , \cite{Charlot2000} dust models, \cite{1983QJRAS..24..267H} grey body dust emission, \cite{Chabrier2003} IMF, \cite{Madau1995} IGM attenuation model and an exponentially declining star formation history model. {The SFR timescale is over the past 100Myr.} 
\\In our analysis we use a modified version of the high\_z version of the code with a lower dust prior, an increased range of available redshifts (beyond z$>10$) and an approximately 10-fold increased model sampling at higher redshifts (z$\sim8$). Dust attenuation is expected to be low at high redshifts due to the early galactic stars being mostly composed of hydrogen and not having yet seeded {a significant metallicity content into the ISM, of which dust is made} {\citep{Shapley2023,Cameron2023}}. {While some dusty galaxies have been discovered in early epochs \citep{Donnan2022}, the original prior is based on local universe observations which are more consistently dusty and so required lowering for improved fits to the higher redshift sample}. The introduction of additional models at high redshift allows us to more finely sample the parameter space and derive a better fit to the given photometry. This was found to be particularly necessary to prevent the highest redshift surveys with low sampling from sharing similar ill-fitting models across many different galaxies.

In order to determine the intrinsic production efficiency of ionizing photons we also remove the IGM absorption component of the SED \underline{after} the other parameters are fit such that their determination is unaffected.
The integration of the $\lambda_{rest}<912$\AA\ region is then used to derive the flux of the Lyman Continuum region. The physical parameters are still modelled using the \cite{Madau1995} prescription and determined by their likelihood distributions. While MAGPHYS simultaneously determines dust and stellar components, we do not report the dust mass or luminosity as our photometric sampling does not constrain the mid-far IR range. 

{\oiii\ ``contaminated" filters also limit the accuracy of the SED fit, due to these bright lines distorting the flux output in their containing filters. For a discussion of the \oiii\ contamination and the method by which we account for this see \ref{subsec:mass}.} Fig \ref{fig:regionplot} breaks down a MAGPHYS SED fit to a ZFOURGE EELG photometry which has been sectioned into wavelength regions of interest {and identifies the ``contaminated" filters}. 

\subsection{Ionizing photon production efficiency log$_{10}(\xi_{ion})$}\label{subsec:sion}
The contribution of a source to reionization is described by the equation 
\begin{equation}
    \dot{n}_{ion} = f_{esc}\times \xi_{ion} \times \rho_{UV}
\end{equation}
where $\dot{n}_{ion}$ is the production rate of ionizing radiation, $f_{esc}$ is the escape fraction of LyC light, log$_{10}(\xi_{ion}/(Hz\  {\rm erg}^{-1}))$ is the production efficiency of ionizing radiation in ${\rm Hz\ erg}^{-1}$ and $\rho_{UV}$ is the comoving UV luminosity density in ${\rm erg\ s^{-1}\ Hz^{-1}\ Mpc^{-3}} $. 


The production efficiency of ionizing radiation \\ log$_{10}(\xi_{ion}/(Hz\  {\rm erg}^{-1}))$ is a key determinant of a galaxy's potential contribution to reionization; describing how much of the integrated spectrum is in the LyC region relative to the non ionizing UV component which represents the young stellar population. Spectroscopically it is determined by the Balmer line luminosities (${\rm H}\alpha$, ${\rm H}\beta$ or ${\rm H}\gamma$) line luminosities in ratio with the luminosity of a non-ionizing UV wavelength such as 1500 \AA\: 
log$_{10}(\xi_{ion}/(Hz\ {\rm erg}^{-1})) = \frac{N_{H^0}}{L_{UV}\times c_{rec}}$
where the $N_{H^0}$ uses the \cite{Leitherer1995} conversion ${\rm N(H^0)}[{\rm s}^{-1}] = \frac{1}{1.36}\times 10^{12}L(H_\alpha)[{\rm erg\ s}^{-1}]$ from luminosity to a production rate of ionizing photons, assuming that the recombination and ionization of the nebula are balanced. $c_{rec} = 2.89$ refers to the case B recombination constant \citep{2006agna.book.....O} for Hydrogen that allows use of either the H$_\alpha$ or H$_\beta$ lines depending on what is available observationally. It should be noted that a dust correction is to be applied to the ${\rm L_{UV}}$ as this is the intrinsic value. 
Photometrically however, we instead use equation 2 from \cite{Wilkins2016a}:
\begin{equation}
    (\xi_{ion}/(Hz\  {\rm erg}^{-1})) = \int^{c/912\AA}_{\infty} L_\nu ({\rm h}\nu)^{-1}/L_\nu(1500{\rm \AA}) {\rm d}\nu
\end{equation}

where $c$ is the speed of light in Angstroms/s and $h$ is the Planck constant in ${\rm erg\ Hz^{-1}}$. This essentially divides the Lyman Continuum region luminosity density by a non ionizing luminosity constant, indicating the relative production of ionizing radiation. This is a value strongly dependent on the stellar population synthesis model used requiring a consistent method for meaningful comparison, so we explore only the MAGPHYS SED code for all samples. The unattenuated (dust corrected) luminosity is used for this calculation as it should reflect the intrinsic production efficiency.

\subsection{UV slope $\beta$}\label{subsec:beta}
The UV slope ($\beta$ where $f_\lambda \propto \lambda^\beta$) is defined as the slope of the SED between rest frame UV wavelengths selected to exclude the LyC /Lyman-$\alpha$ features and to set an upper limit still considered to be in the UV region. It is commonly used as an indicator of dust attenuation and the age of the stellar population \citep{Wilkins2013,Williams2018,Reddy2018,Nanayakkara2022} {which are particularly important for the analogue analysis in this work}.

\begin{figure}
    \centering
    \includegraphics[width=\linewidth, height=0.40\textheight]{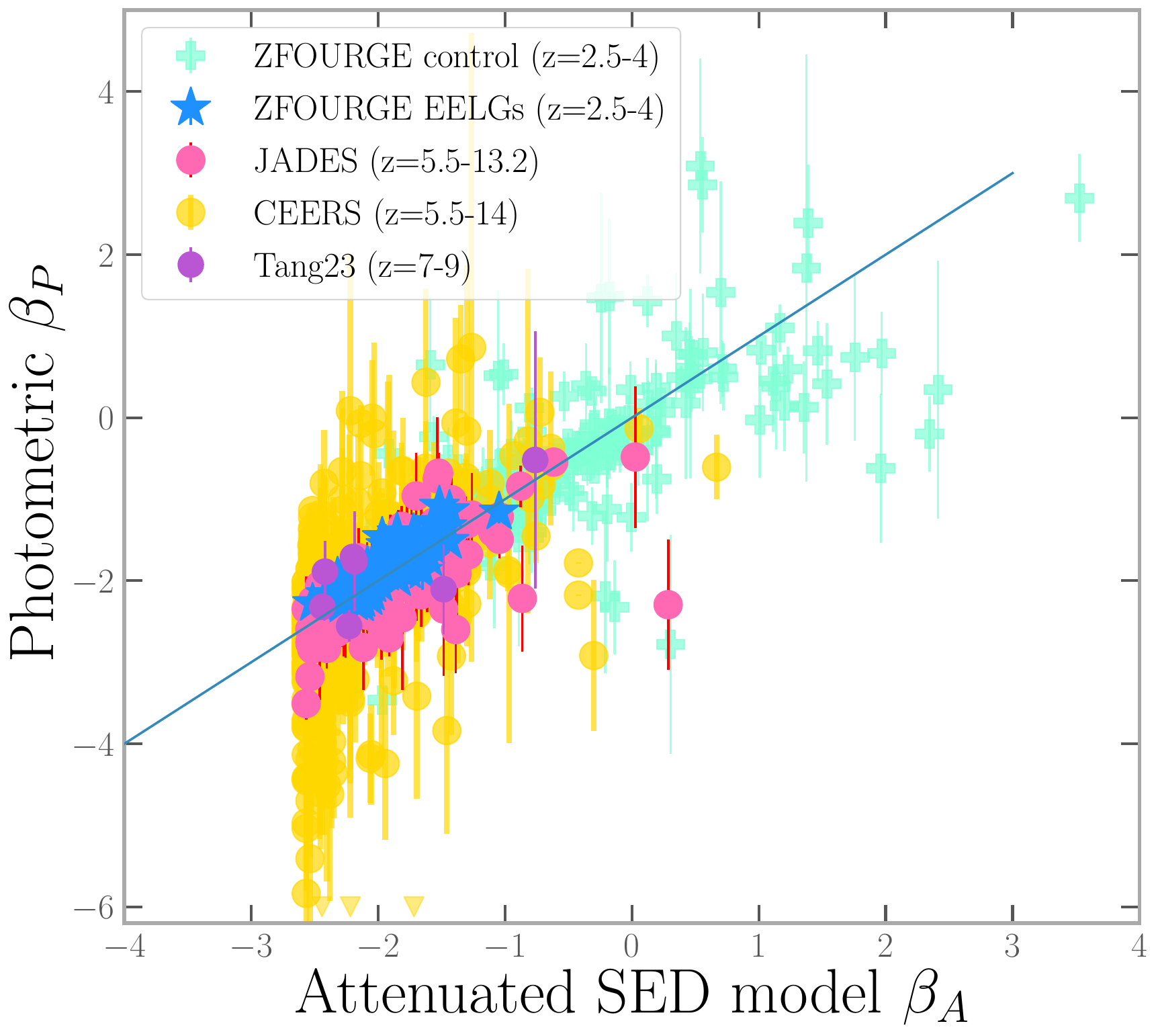}
    \caption{Attenuated model ($\beta_a$ vs direct photometry $\beta_p$ UV slopes for each subsample, with the {ZFOURGE }EELGs as blue stars, {ZFOURGE control in cyan pluses, and the EoR samples in pink (JADES), gold (CEERS) and purple (Tang23) circles.} {The 1 to 1 line is shown in blue.} General agreement is found with the exception of the {CEERS sample around the }$\beta_p < -2.6$ region, where the model attempts to constrain the relative number of blue stars {that could have formed in the galaxy's lifetime, limiting the final ``blueness" of the slope}.}
    \label{fig:bpvsba}
\end{figure}

We further define two versions of the UV slope $\beta$ parameter that reflect each the blueness of the direct (not dust corrected) photometry $\beta_P$ and that of the {model derived }attenuated SED $\beta_A$ between rest wavelengths 1300-2600 ${\rm \AA}$ (the orange highlighted region of Fig \ref{fig:regionplot}) {(see \cite{Calzetti1994} for a discussion of the chosen wavelength range)}. These are determined using the scipy.optimize.curve\_fit package and under the constraint that at least 3 filters between this wavelength range have a non-zero flux.

The direct photometry is less commonly considered in these analyses \citep{Rogers2013} as the model dependence of the redshift determination creates a degeneracy with the star formation history model parameters that determine the attenuated UV fit. However, our data includes both spectroscopic redshifts and {highly accurate photometric redshifts ($<$2\% errors), therefore we can estimate the attenuated UV slope while considering the errors in the observed photometry without relying on the underlying models.} A discussion on the value of this parameter in contrast to the model derived attenuated slope can be found in \cite{Dunlop2013}. {Fig \ref{fig:bpvsba} highlights the key differences between these. }

\begin{figure}[h!]
    \centering
    \includegraphics[width=1.05\linewidth, height=0.40\textheight]{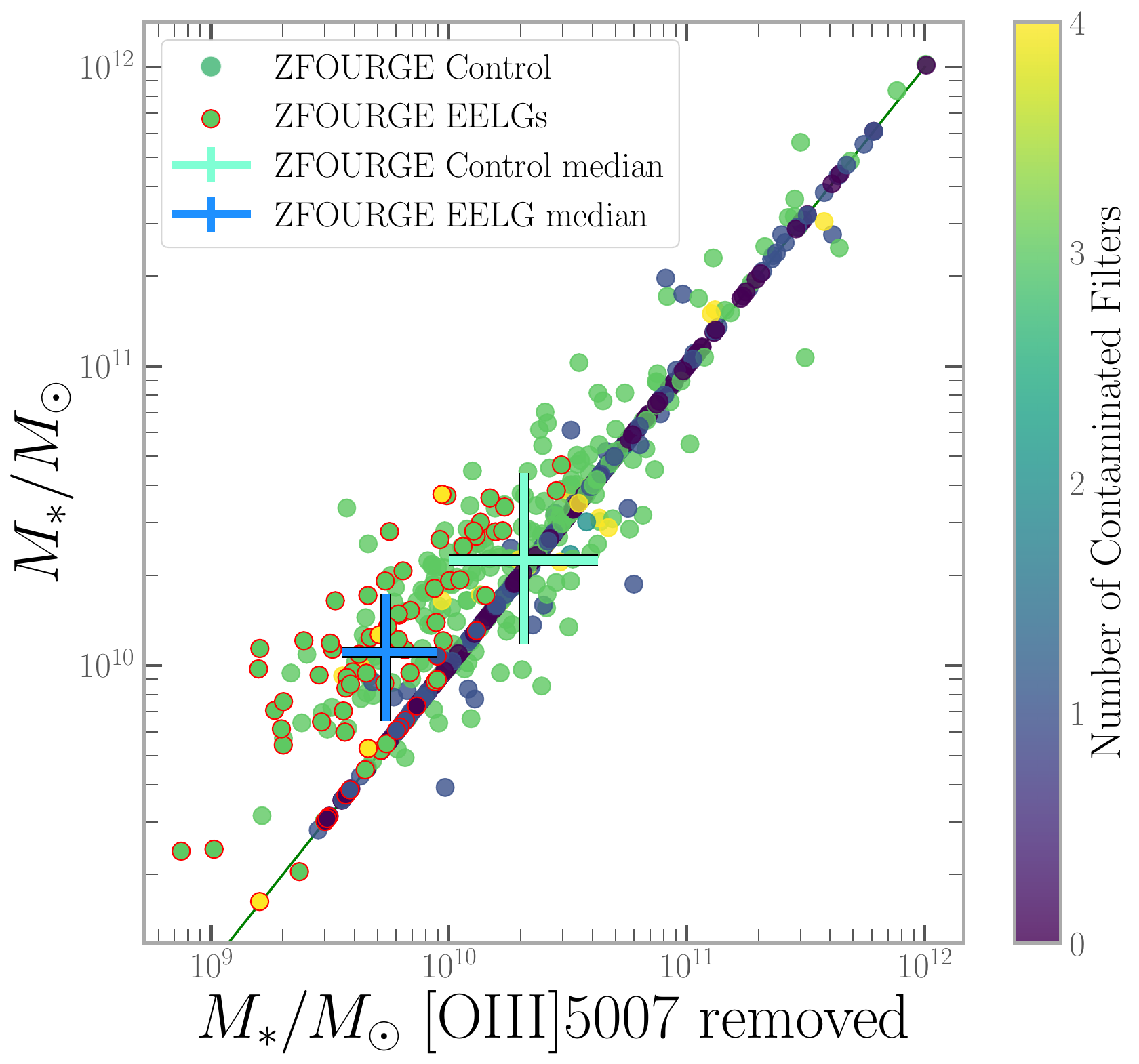}
    \caption{Stellar Mass with and without inclusion of the \oiii\ contaminated filters for the ZFOURGE control and {ZFOURGE }EELG samples. We also note a trend that $2.9<z<3.5$ galaxies tend to be more affected by the contamination and our EELG subsample is on the higher end of the discrepancy. This is due to the tendency for the \oiii\, line to fall into multiple filters at this redshift, and this is represented by the colour {bar}.}
    \label{fig:masscontam}
\end{figure}
\begin{table}[htbp]
    \centering
    \caption{Parameter medians with 25-75 percentiles {for the ZFOURGE control, ZFOURGE EELGs, JADES and CEERS samples}}
    \label{tab:medians}
    \begin{tabular}{lccccp{2cm}}
        \toprule
        & ZFOURGE & EELGs & JADES & CEERS \\
        \midrule
        $log_{10}(M_*$/$M_{\odot})$ &$10.32^{0.32}_{-0.29}$ &$9.75^{0.20}_{-0.28}$ &$8.43^{0.37}_{-0.28}$ &$8.71^{+0.37}_{-0.36}$  \\
        $log_{10}$(sSFR/$yr^{-1}$) &$-8.48_{-0.37}^{+0.35}$ &$-8.28_{-0.23}^{+0.24}$ & $-7.99_{-0.07}^{+0.12}$&$-8.17_{-0.18}^{+0.21}$ \\
        $A_V$ &$0.57^{0.30}_{-0.20}$ &$0.25^{0.12}_{-0.14}$ &$0.22^{0.42}_{-0.16}$ &$0.08^{0.09}_{-0.04}$ \\
        $\beta_P$ &$-1.27^{0.31}_{-0.28}$ &$-1.83^{0.21}_{-0.15}$ &$-2.02^{0.38}_{-0.31}$&$-2.20^{0.50}_{-0.49}$ \\
        log$_{10}(\xi_{ion}/{(\rm Hz erg}^{-1}))$ &$25.14^{0.06}_{-0.04}$ & $25.16^{0.06}_{-0.05}$&$25.13^{0.06}_{-0.05}$ &$25.18^{0.06}_{-0.07}$ \\
        \bottomrule
    \end{tabular}
\end{table}
\begin{figure}[h]\
    \centering
    \includegraphics[width=\linewidth, height=0.40\textheight]{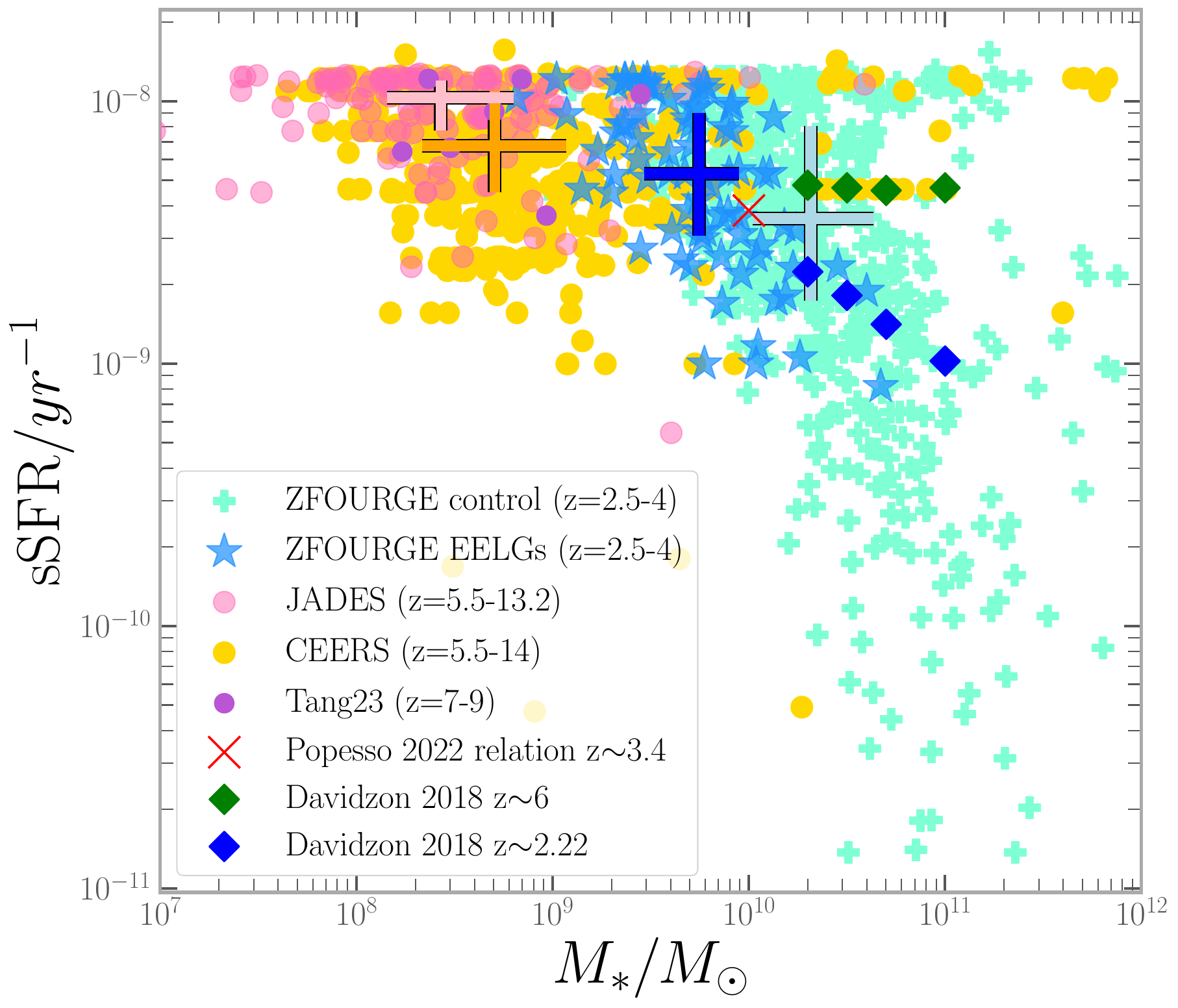}
    \caption{Specific star formation rate vs Stellar Mass plot for each of our samples. Colour-scheme is the same as Figure \ref{fig:bpvsba} The median values and interquartile ranges are indicated by the coloured errorbars with orange:CEERS, pink:JADES, dark blue :{ZFOURGE }EELGs, teal:ZFOURGE {control}. The upper boundary is caused by the chosen SFR timescale. {\cite{Popesso2022} z-sSFR-$M_*$ relation selected at $z\sim3.4$ and $M_*/M_\odot=10^{10}$ shown as a red cross, and \cite{Davidzon2018} at $z=2.22$ and $z=6$ shown as a blue and green diamonds for comparison.}}
    \label{fig:mainseq}
\end{figure}

\begin{figure}[h]
    \centering
    \includegraphics[width=\linewidth, height=0.40\textheight]{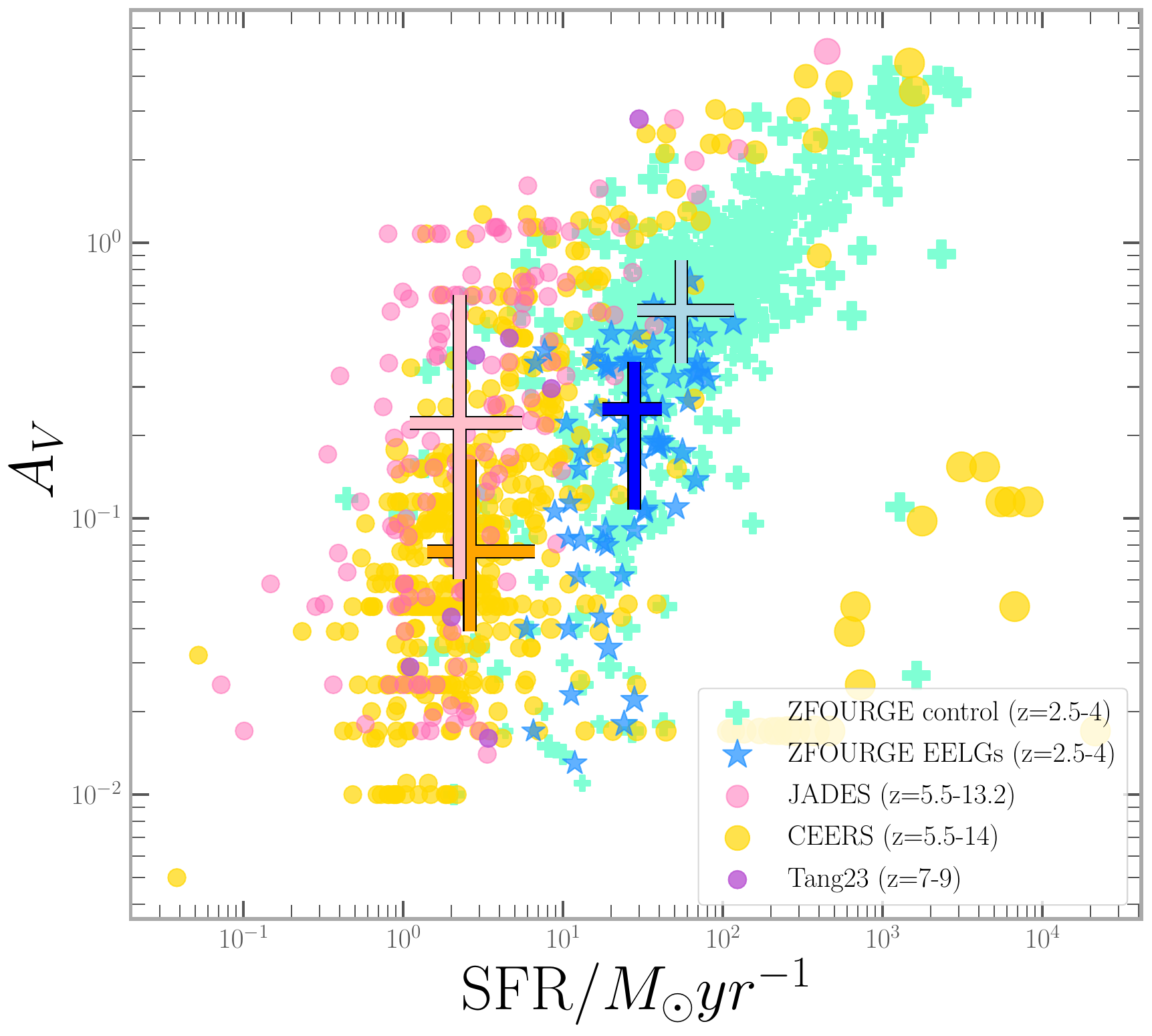}
    \caption{Star formation rate vs Dust attenuation plot for each of our samples. Marker sizes correlate to the Stellar Mass of the galaxy, with larger markers representing more massive galaxies consistent across all samples. Colour scheme represents the same samples as Figure \ref{fig:bpvsba}. median values and interquartile ranges are indicated by the coloured errorbars following the same colour-scheme as Figure \ref{fig:mainseq}}
    \label{fig:sfratten}
\end{figure}

\begin{figure*}[t]
    \centering
    \includegraphics[width=\textwidth, height=0.24\textheight]{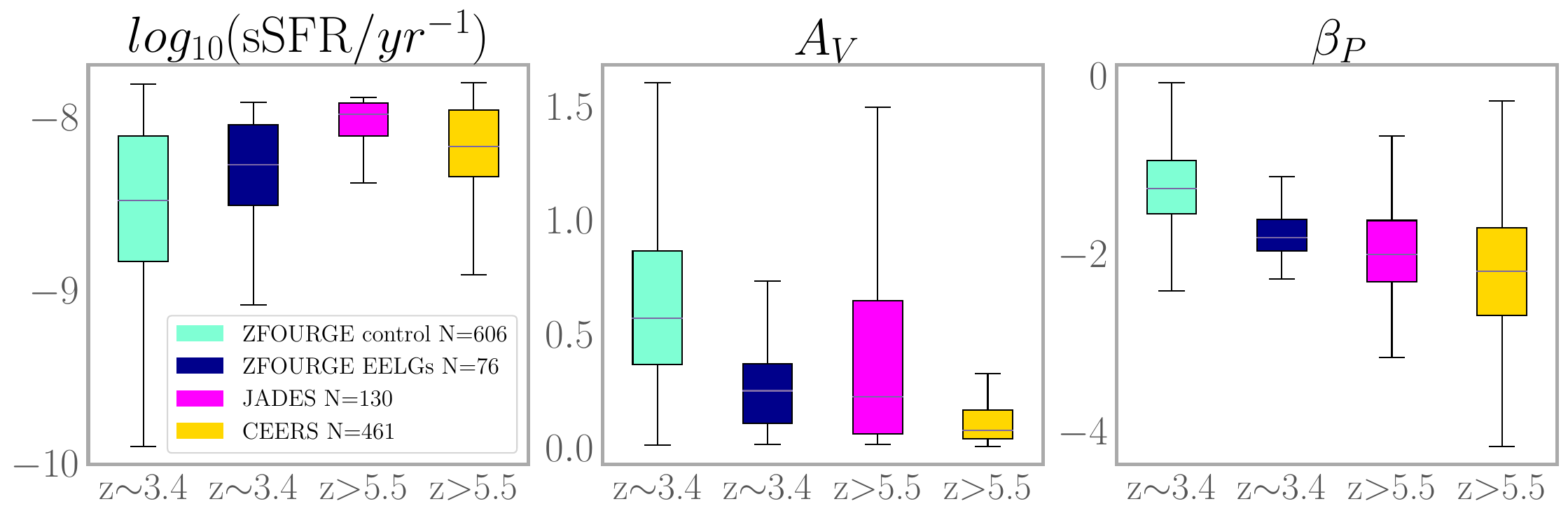}
    \includegraphics[width=\textwidth, height=0.24\textheight]{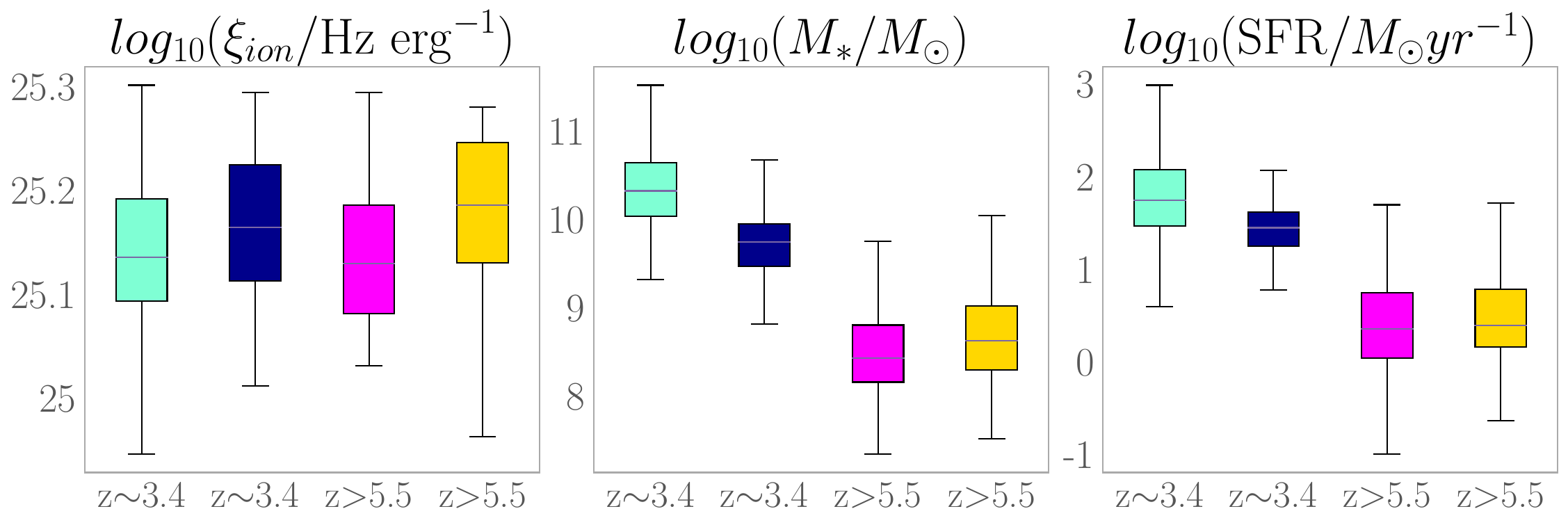}
    
    \caption{MAGPHYS derived physical parameters for each of the samples following the same colour scheme as previous figures. Median values and correlations can be found in Tables \ref{tab:medians},\ref{tab:correlations}.}
    \label{fig:beta}
\end{figure*}
\begin{figure*}[t]
    \centering
    \includegraphics[width=\textwidth, height=0.24\textheight]{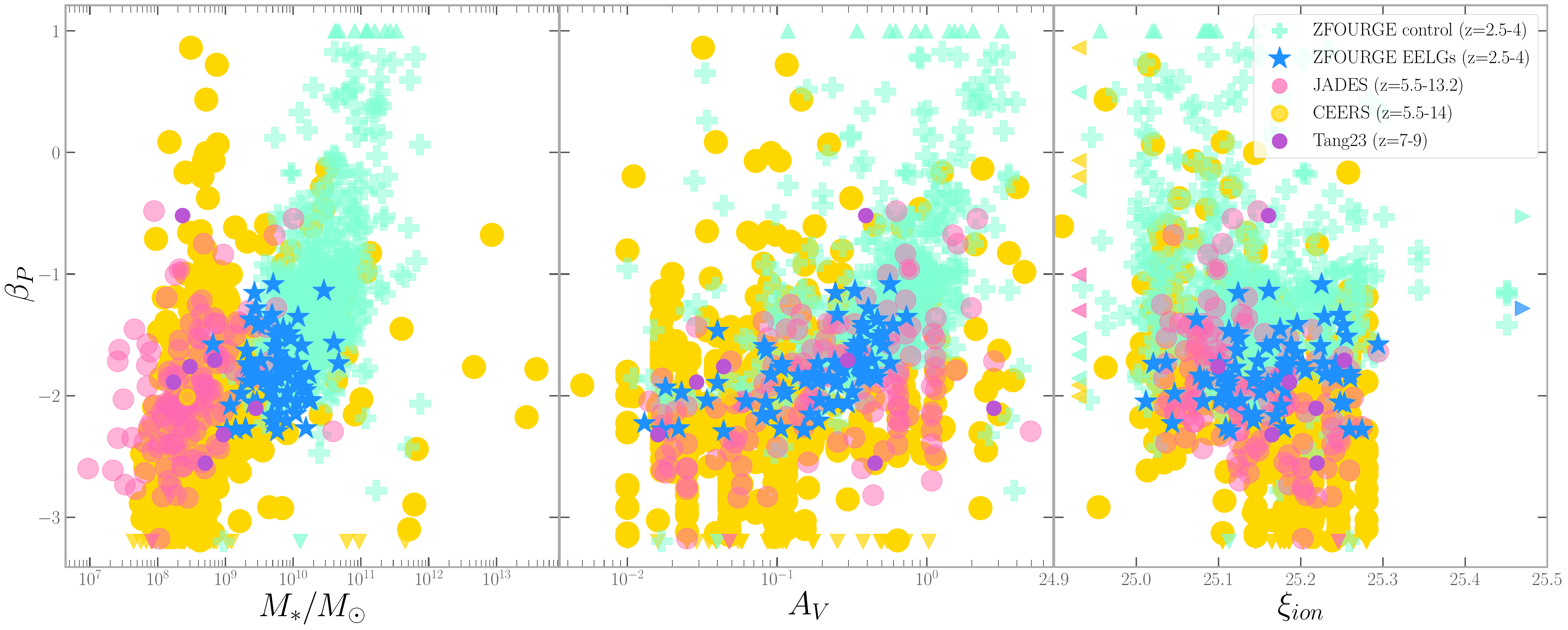}
    \includegraphics[width=\textwidth, height=0.24\textheight]{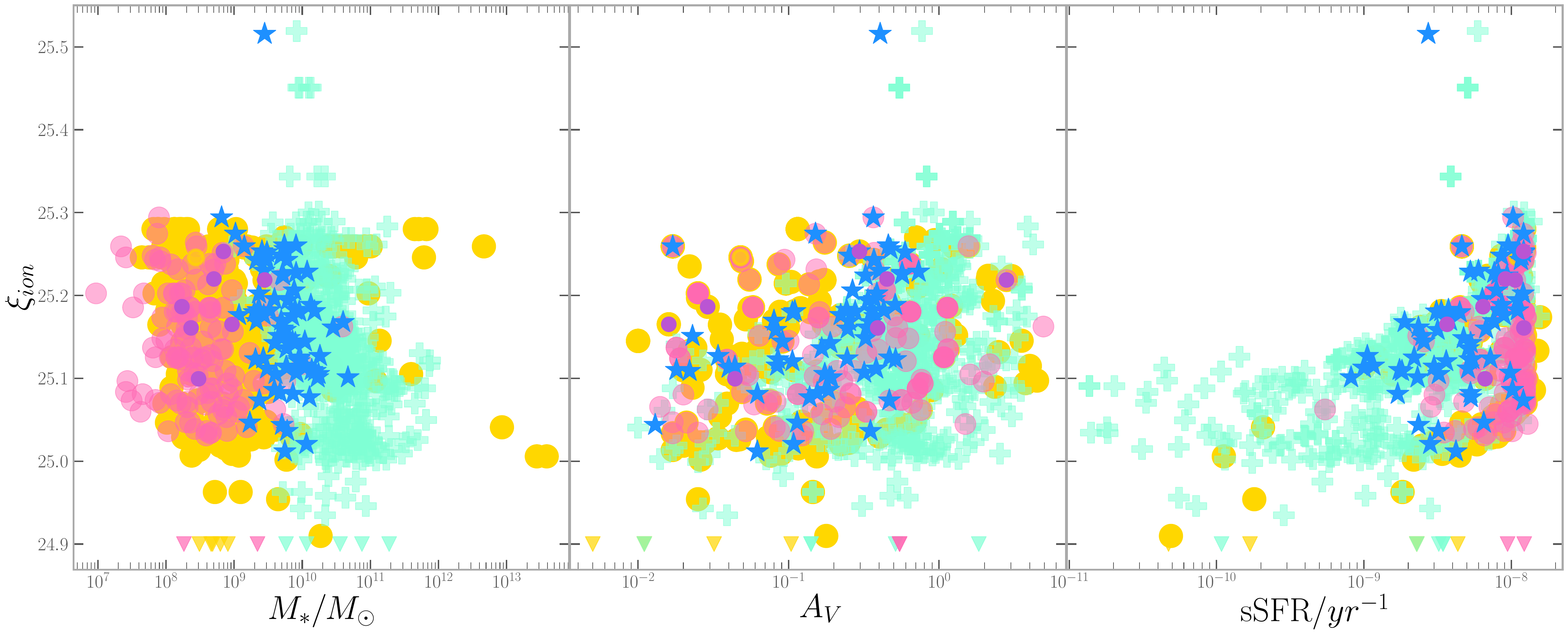}
    \caption{MAGPHYS derived physical parameters vs  $\beta_P$ (uncorrected for dust, top panel) vs photometric log$_{10}(\xi_{ion}/(Hz\  {\rm erg}^{-1}))$ (dust corrected \citep{Charlot2000}, bottom panel) for each of the samples following the same colour scheme as previous figures. Median values and correlations can be found in Tables \ref{tab:medians},\ref{tab:correlations}. Model limitations manifest as ``walls" in $\xi_ion$ and sSFR.}
    \label{fig:sion}
\end{figure*}
The depth of the CEERS data as well as the limited photometric sampling could significantly distort the value derived from the direct photometric method, however, the model slope is not a solution to this. {SED models are typically optimised for typical star forming galaxies at $z<6$ and hence the modelled UV slope may not be representative of the observed photometry for extreme cases. The model-independent UV slope is robust as it is estimated directly from the photometry and reflects the associated errors. Thus, it provides a consistent, comparative tool between the analogues and EoR galaxies.}  

\subsection{Mass correction in EELGs \oiii\ contamination}\label{subsec:mass}

The \oiii\ emission line flux poses as a significant contaminant to the photometric filters in which it resides if not taken into consideration \citep{Forrest2018}. {The filter containing the emission line adds this to the total measured continuum flux, which is then read by the SED fitting code to determine physical parameters}. {Significantly bright} emission lines raise {the measured continuum }flux above the {real} continuum, resulting in a {flux excess} and causing the physical parameters associated with that portion of the SED to be {distorted if not correctly accounted for, as is the case with the current version of MAGPHYS}.
\\Our sample selection is based on the \oiii\ EW, though we entirely remove the \oiii\ contaminated filters while fitting the SED of galaxies with MAGPHYS. Figure \ref{fig:regionplot} shows the effects of this on {an example} SED while Figure \ref{fig:masscontam} shows the effect of {contamination} on the stellar mass of the ZFOURGE galaxies and the EELG subsample. {We note} that EELG galaxies around $z\sim 3.5$ are most significantly affected by this. The mass determination is dependent largely on the optical region of the SED and the heightened \oiii\ {flux} in this subsample is expected to contribute to the {four filters in the K}-band {($\sim 2$ micron)} more-so than in most galaxies.
The inclusion of emission lines in the SED fitting models themselves is another way of accounting for this effect which is employed in other SED fitting codes (e.g. \cite{Chevallard2016}). The same removal method was applied to the JADES sample but not the CEERS sample, as its further limited photometric coverage would significantly reduce the model reliability. While this {may} result in a moderately overestimated mass {in the CEERS sample due to the expected high average \oiii\ EW}, this alone would not account for the extreme values of some of these galaxies which require better photometric constraints {(see section \ref{subsec:main sequence} for further discussion)}.

\begin{table}[htbp]
    \centering
    \caption{Spearman rank correlations for each subsample {(ZFOURGE control, ZFOURGE EELGs, JADES and CEERS)} and the combined dataset (Total). Blank spaces reflect correlations above a significance factor of 0.05.}
    \label{tab:correlations}
    \begin{tabular}{lccccc}
        \toprule
        & ZFOURGE & EELGs & JADES & CEERS & Total\\
        \midrule
        \multicolumn{6}{c}{{Correlation with $\beta_P$}} \\ 
        \midrule
        $log_{10}(M_*$/$M_{\odot})$ &0.48 &... &0.45 &0.45 &0.67 \\
        $log_{10}$(sSFR/$yr^{-1})$ &-0.08 &... &0.32 & -0.12&-0.24 \\
        $A_V$ &0.52 &0.63 &0.44 &0.21 &0.58 \\
        
        \midrule
        \multicolumn{6}{c}{{Correlation with log$_{10}(\xi_{ion}/(Hz\  {\rm erg}^{-1}))$}} \\
        \midrule
        $log_{10}(M_*$/$M_{\odot})$ &-0.48 &-0.36 &-0.24 &-0.36 &-0.32 \\
        $log_{10}$(sSFR/$yr^{-1})$ &0.73 &0.57 &... &0.46 &0.55 \\
        $A_V$ &0.32 &0.53 &... &... &... \\
        $\beta_P$ &-0.18 & ...&-0.42 & -0.44&-0.33 \\
        \midrule
        \multicolumn{6}{c}{{Correlation with $A_V$}} \\
        \midrule
        $log_{10}$(SFR/$M_{\odot}$/yr) &0.70 &0.48 &0.56 &0.45 &0.68 \\
        $log_{10}$(sSFR/$yr^{-1})$ &0.39 &0.59 &0.51 &0.29 &0.16 \\
        \bottomrule
    \end{tabular}
\end{table}

\section{Results and Discussion} \label{sec:results}

{The physical parameters relating to the stellar population and dust attenuation are analysed in the {ZFOURGE }EELG sample and compared to the control samples within the EoR (JADES, CEERS, Tang23) and at similar redshift {(ZFOURGE control)}.}
{The comparisons selected explore the consistency with which the $2.5<z<4$ \oiii\ analogues match the physical parameters and internal processes of EoR galaxies.}

\subsection{Main Sequence}\label{subsec:main sequence}
We {find} that the {ZFOURGE }EELG subsample is consistently on the lower mass, higher specific star formation rate end of this parameter space when compared to the ZFOURGE control sample (Fig \ref{fig:mainseq}). While {ZFOURGE }EELG stellar masses are generally an order of magnitude above the EoR counterparts, the sSFR median and 25-75 quartile region {is} comparable in the {ZFOURGE }EELG subsample and the JWST survey samples between redshifts 5.5-14 (Table \ref{tab:medians}). Our ZFOURGE {control} sample at $z\sim3.4$ lies close to the stellar mass-redshift-sSFR relation developed by \cite{Popesso2022} using 27 other studies {between $0<z<6$ and $8.5<log_{10}(M_*)<11.5$}. The {sSFR of the} {ZFOURGE }EELG sample, however, is 0.2 dex above this relation \citep{Popesso2019,Leslie2020} and more closely related to the semi-empirical samples at $z\sim6$ \citep{Grazian2015,Davidzon2018}. This reflects the strength of the \oiii\ EW selection technique at low redshift in collating the most highly star forming galaxies.
We note the appearance of striations visible in the CEERS sample. This is due to the {relatively lower SNR and fewer photometric filters across the spectrum}; preventing the code from fitting models to the more intricate variations in the continuum and thus recycling similar models. This effect continues through the other physical parameters to which the CEERS sample is fit and should be considered accordingly.

{The stellar mass and star formation rates of both the ZFOURGE control and EELG samples are uniformly higher than that of the EoR samples, with the EELG sample being slightly less massive and less star forming. While the {ZFOURGE }EELG mass is higher than the EoR samples, it is still $\sim 25\%$ less massive than the average population at $z\sim3$. The emission of LyC radiation is also not directly dependent on this parameter, so the functionality as an analogue is maintained. The heightened SFR is related to the emission, however, when normalizing by mass (sSFR) {we will see that} the discrepancy between EELG analogues and EoR galaxies is minimised.}

\subsection{Dust Extinction Star Formation relation}\label{subsec:dust}
Dust and star formation are inextricably linked galactic processes. Dust cools and catalyses molecular gas collapse while stellar death synthesizes more dust and transports it through the galaxy \citep{Popping2017}. {The correlation of these properties across different epochs could be indicative of their evolving relationship as galaxies age and their dust content increases. Studies such as \cite{Zahid2012} link the correlation between $A_V$ and SFR to the processes that quench star formation. A positive Spearman rank correlation between {the dust extinction and star formation rate is observed by \cite{Li2019} in nearby galaxies}, while \cite{Zahid2012} found a mass dependence on the correlation. They found more massive galaxies to be positively correlated while finding an anticorrelation below $10^{10}M_\odot$\, when the quiescent high mass sample was removed, indicative of differing internal processes governing this relation between the low and high mass samples. Interestingly, the Balmer decrement ($A_V$ tracer) stellar mass relation shows minimal evolution with redshift for $z<2$ sources \citep{Battisti2022}}.

{For galaxies to be analogues of EoR galaxies, they should exhibit similar correlations across various physical parameters, owing to similar internal physical environments. Here we compare our findings across different epochs with our self-consistent, single SED fitting method to determine if the \oiii\ selected {ZFOURGE }EELG sample has a similar relationship between these properties to EoR galaxies.} {We find that} {ZFOURGE }EELGs have a consistently lower dust extinction value when compared to the ZFOURGE control sample Table \ref{tab:medians}. The median value is half that of the ZFOURGE control and is well within the 25-75th percentiles of the EoR JADES sample.

We find a positive correlation between SFR and dust attenuation across all our samples. The correlation appears strongest in the ZFOURGE control sample, {which tend to have higher $A_V$ and SFR than the others (See Table \ref{tab:correlations}). This finding is similar to \cite{Sakurai2013} who also find a stronger correlation between $A_V$ and SFR for galaxies with SFR $>20 M_\odot/yr$}. {However, the exact values of our correlations may be biased by the intrinsic scatter and parameter ranges. For example, the weaker correlations observed in {ZFOURGE }EELGs and EoR (JADES, CEERS) samples could be due to the lack of dusty, low star-forming galaxies in {the samples}.}
 
{Even after accounting for stellar mass by comparing sSFR and $A_V$, we still find significant positive correlations across the four samples.} The weaker correlations in the ZFOURGE and CEERS samples could be due to their relatively large parameter ranges compared to the {ZFOURGE }EELG and JADES samples. {The similar} correlation in the EELG and JADES samples in contrast to the ZFOURGE control suggests that the physical characteristics of {ZFOURGE }EELGs are more comparable to the z$>5.5$ JADES sample than to the control z$\sim3.4$ sample.


\subsection{UV slope ($\beta_p$) relations}\label{subsec:slope}
We compare the different samples parameter space relations in Fig \ref{fig:beta} and identify the correlations between each subsample in Table \ref{tab:correlations}.  
The UV slope {interquartile range} of the {ZFOURGE} EELGs matches with the upper quartile of the JADES sample and is well below the ZFOURGE control (Figure \ref{fig:beta} top panel). 
{This identification of the {ZFOURGE }EELG sample as being similarly blue and dust free as the direct EoR samples} while having a higher specific star formation rate than the ZFOURGE control sample (see Table \ref{tab:medians}) { indicates a strong potential for these to be good EoR analogues.} 

The $\beta_P$ vs $M_*$ relation (Fig \ref{fig:sion}) appears to be discontinuous between the low and high redshift samples. The observed correlation between stellar mass and slope agrees with the conclusions of \cite{Pannella2009} and with the consensus that increasing stellar mass correlates with increasing dust attenuation and therefore a redder ({observed}) slope \citep{Bouwens2016}. This is supported by the lower dust extinction (Table \ref{tab:medians}) of the high redshift samples, as well as explains the tight parameter space occupancy of the {ZFOURGE }EELG sample in this figure. The correlation between UV slope and stellar mass is not present in the {ZFOURGE }EELG sample (Table \ref{tab:correlations}) as it occupies a very constrained portion of the parameter space compared to the other samples (1.8 dex in stellar mass for {ZFOURGE }EELGs vs 3.6 dex for JADES for example). The UV slope appears to be sensitive to the mass with a strong overall correlation of 0.67 and strong correlations with the EoR and control samples. This suggests that the {ZFOURGE }EELG sample is consistently blue despite the mass being an order of magnitude greater than the EoR JADES median. 

The {correlation between the UV slope and dust extinction} is readily identifiable, being strongest in the ZFOURGE {control} and EELG samples, which is similar to the conclusion drawn in \cite{Wilkins2013}. The parameter space occupied by the {ZFOURGE }EELGs is the same as that of the JADES and Tang23 samples which could indicate the `blueness' of a galaxy being attributed to the absence of attenuating dust in the model, though better dust constraints are required. 

We also note {the negative correlation of the \\ $\beta_P$ with} $\xi_{ion}$ plot (Fig \ref{fig:sion}). The $\xi_{ion}$ does not evolve significantly between the different redshift samples, however it does appear to negatively correlate with UV slope in both the EoR samples. {This suggests that at least for the EoR samples, bluer galaxies tend to have a higher ionizing to non ionizing photon ratio though further investigation of the model dependence of the $\xi_{ion}$ parameter are necessary.}


\subsection{Ionizing photon production }\label{subsec:xi_ion}
This section explores the correlations between the ionizing photon production efficiency $\xi_{ion}$ and the physical parameters determined by our model (Figure \ref{fig:sion}). We refer to Table \ref{tab:correlations} for the correlation coefficients and their significance.

The $\xi_{ion}$ {anticorrelates} with stellar mass in the low redshift samples (-0.48 ZFOURGE control, -0.36 {ZFOURGE }EELGs), suggesting that a lower mass corresponds to a higher production efficiency, though this is not clear from its visual {appearance} (Fig \ref{fig:sion}). This weak dependence on stellar mass agrees with the findings of \cite{Emami2020} and \cite{Shivaei2018} for this mass range ($10^{7}-10^{11} M_\odot$) at a similar redshift range to our study, as well as with \cite{Lam2019} for the EoR samples. This weak anticorrelation with stellar mass potentially implicates smaller systems at early epochs as the powerhouses of ionizing photon production.

The production efficiency correlates most strongly with the sSFR in the ZFOURGE {control} and {its} EELG {subsample}. The assertion that a high specific star formation rate would be reflected in higher LyC emission is known \citep{Castellano2023} between $2<z<5$. While a significant correlation could not be determined for the JADES sample due to the small dynamical range of sSFR values, the CEERS sample does reveal a similarly strong relationship in this parameter space. 

{A notable correlation between $\xi_{ion}$ and dust extinction is found only in the {ZFOURGE }EELG sample. The result suggests a relationship between the dust within {\oiii\ selected} galaxies and their efficiency of LyC photon production. This is likely due to the underlying degeneracy correlating $A_V$, age and SFR. Young galaxies with emerging O star populations haven't seeded as much dust when compared to their older counterparts, and it is these massive stars that are the driving force for the strong \oiii\ emission lines in the {ZFOURGE }EELG sample. }

\section{Summary and Conclusion}\label{sec:summary}
This paper uses SED modelling to derive the physical parameters of EELGs {with analogous \oiii\ emissions to EoR galaxies} at $z\sim3$ and compares them to both a control sample as well as {the EoR} galaxies observed at $z>5.5$ with JWST. In this section we summarize the findings of our analysis and the potential avenues further research can take.\\

{1. The combined high sSFR (-8.28$yr^{-1}$), comparatively low stellar mass (9.75$M_\odot$), low dust extinction ($A_V$= 0.25) and blue UV slope ($\beta_P$= -1.84) {suggest that the \oiii\ selection technique is a good analogue selector. The high SFR and low dust are critical for their analogue status as they must produce copious ionizing radiation that is not attenuated by dust. We need to further investigate the escape of LyC light by studying the $f_{esc}$ to complete the picture}.}\\
2. We confirm the strength of the \oiii\ selection technique for finding {EoR analogues} with low{er} stellar mass, high relative star formation {rates} for our sample between $2.5<z<4$. These galaxies are an order of magnitude higher in mass than their EoR counterparts but maintain similar sSFR {despite this}.\\
3. {We find that the \oiii\ selected EELGs have a more similar correlation between their dust and star formation parameters to EoR galaxies than to the control sample between $2.5<z<4$, suggestive of similar internal processes relating to the dust distribution and star formation between the analogues and EoR galaxies.}\\
4. We find a correlation between the UV slope and dust extinction/attenuation which is strongest in our low redshift samples. We also see a similar correlation with the stellar mass, particularly in our control sample and the CEERS data. Experiencing similar attenuation between these samples indicates low dust obscuration however this is only supported by UV photometry. Further studies using deep ALMA observations are required to determine if the steep UV slopes are due to the stellar population or an absence of dust.\\
5. We find that the median $\beta_P$ of our {ZFOURGE }EELGs matches more {closely} with the higher redshift counterparts than with the $2.5<z<4$ control sample, suggesting that our selection technique has recovered galaxies with a similar starbursty nature as the EoR galaxies.\\
6.{We find the most significant correlation of log$_{10}(\xi_{ion}/(Hz\  {\rm erg}^{-1}))$ to be with its sSFR which coincides with the findings of \cite{Castellano2023} for bright $M_UV\leq 20$, $2<z<5$ galaxies. This correlation appears the strongest with our low redshift samples{, likely due to the wider parameter space}.} \\


\begin{acknowledgement}
This research was partly supported by the Australian Research Council Centre of Excellence for All Sky Astrophysics in 3 Dimensions (ASTRO 3D), through project number CE170100013. CMT was supported by an ARC Future Fellowship under grant FT180100321.
\\The International Centre for Radio Astronomy Research (ICRAR) is a Joint Venture of Curtin University and The University of Western Australia, funded by the Western Australian State government. AH acknowledge support from the ERC Grant FIRST- LIGHT and Slovenian national research agency ARRS through grants N1-0238 and P1-0188.

\,RJ would like to thank Jessica E Thorne for helpful discussions.
\end{acknowledgement} 

\newpage
\bibliography{example}



\end{document}